\documentclass[a4paper]{jpconf}
\usepackage{amssymb}
\usepackage{graphicx}

\newcommand{\be}{\begin{equation}}
\newcommand{\ee}{\end{equation}}

\newcommand{\bea}{\begin{eqnarray}}
\newcommand{\eea}{\end{eqnarray}}

\begin{document}
\title{The relationship between trading volumes, number of transactions, and stock volatility in GARCH models}

\author{Tetsuya Takaishi$^{1}$ and Ting Ting Chen$^{2}$ }
\address{$^{1}$Hiroshima University of Economics, Hiroshima 731-0192, JAPAN}
\address{$^{2}$Faculty of Integrated Arts and Sciences, Hiroshima University, Higashi-Hiroshima 739-8521, Japan}

\ead{$^{1}$tt-taka@hue.ac.jp}

\begin{abstract}
We examine the relationship between trading volumes, number of transactions, and volatility using daily stock data of the Tokyo Stock Exchange. Following the mixture of distributions hypothesis, we use trading volumes and the number of transactions as proxy for the rate of information arrivals affecting stock volatility. The impact of trading volumes or number of transactions on volatility is measured using the generalized autoregressive conditional heteroscedasticity (GARCH) model. We find that the GARCH effects, that is, persistence of volatility, is not always removed by adding trading volumes or number of transactions, indicating that trading volumes and number of transactions do not adequately represent the rate of information arrivals. 
\end{abstract}
\vspace{-10mm}

\section{Introduction}
The empirical properties of asset returns have been intensively studied, and some universal properties are classified as "stylized facts" \cite{Cont}. The notable stylized facts include (1) no significant autocorrelation in returns, (2) long autocorrelation in absolute returns, (3) fat-tailed return distributions, and (4) volatility clustering.
The return dynamics explaining these stylized facts have been the subject of numerous studies. Assuming that the return dynamics can be described by a Gaussian random walk with time-varying volatility, one possible explanation is that $r_t=\sigma_t \epsilon_t$, where $r_t$ is a return, $\sigma_t^2$ represents volatility, and $\epsilon_t$ is a random variable from $N(0,1)$ at time $t$. Several studies have verified this assumption \cite{RV1}--\cite{RV7} by examining whether $r_t/\sigma_t$ is consistent with the random variable $\sim N(0,1)$.

Yet another unresolved issue relates to volatility dynamics. Under the mixture of distributions hypothesis (MDH) proposed by Clark \cite{Clark},
volatility dynamics is related to the rate of information arrivals to the market. Since the rate of information arrivals is latent and unobservable, Clark used trading volume as a proxy for the rate of information arrivals. Empirical evidence indicates the existence of a contemporaneous correlation between volatility and trading volume; see, for example, \cite{Karpoff}.

On the other hand, the dynamic behavior of volatility is well captured by the autoregressive conditional heteroscedasticity (ARCH) model \cite{ARCH}
and its extension, the generalized ARCH (GARCH) model \cite{GARCH}. 
In particular, the GARCH model successfully captures the persistence of volatility variation, referred to as GARCH effects. In the GARCH model, the volatility process is described by a function of past volatilities and returns. The MDH also implies that the volatility process is described by a function of trading volume. 
Lamourex and Lastraps \cite{LL} inserted trading volume into the GARCH process by using individual stocks in the US market 
and found that the GARCH effects disappear, supporting the MDH.
Some subsequent studies, for example, \cite{P1,P2,P3,P4}, also support their finding, that is, that the inclusion of trading volume in the GARCH model reduces the GARCH effects.
On the other hand, some other studies, such as \cite{N1,N2,N3,N4,N5}, report that the inclusion of trading volume in the GARCH model does not completely remove the GARCH effects; thus, the MDH is not supported.

In order to elaborate the volatility dynamics, we examine the relationship between trading volume and stock volatility by
using the daily stock data of the Tokyo Stock Exchange from  June 3, 2006, to December 30, 2009.
Specifically, by including trading volumes into the GARCH process, we can infer the GARCH parameters and 
examine whether the GARCH effects can be explained by trading volume.
We also use the number of transactions as a proxy for the rate of information arrivals and examine its effect on GARCH volatility. 

\section{GARCH Test}
We focus on the GARCH(1,1) model \cite{GARCH} described by $r_t =\sigma_t\epsilon_t$, and
\be
\sigma_t^2 = \omega + \alpha r_{t-1}^2 +\beta \sigma_{t-1}^2,
\ee
where $\alpha,\beta$ and $\omega$ are the GARCH parameters to be determined. The magnitude of persistence of volatility, that is, the GARCH effects, is measured by $\alpha+\beta$, and
for high persistence of volatility, we observe that $\alpha+\beta$ is close to 1.
The effect of trading volume or the number of transactions is examined by adding a term to the GARCH process, as
\be
\sigma_t^2 = \omega + \alpha r_{t-1}^2 +\beta \sigma_{t-1}^2 +\gamma N_t,
\ee
where $N_t$ stands for either the trading volume or number of transactions at time $t$.
We infer the GARCH parameters by the Bayesian inference conducted 
using the Markov Chain Monte Carlo (MCMC) method \cite{Takaishi1}--\cite{Takaishi6}.

\begin{figure}
\vspace{-5mm}
\centering
\includegraphics[height=6.7cm,width=13cm]{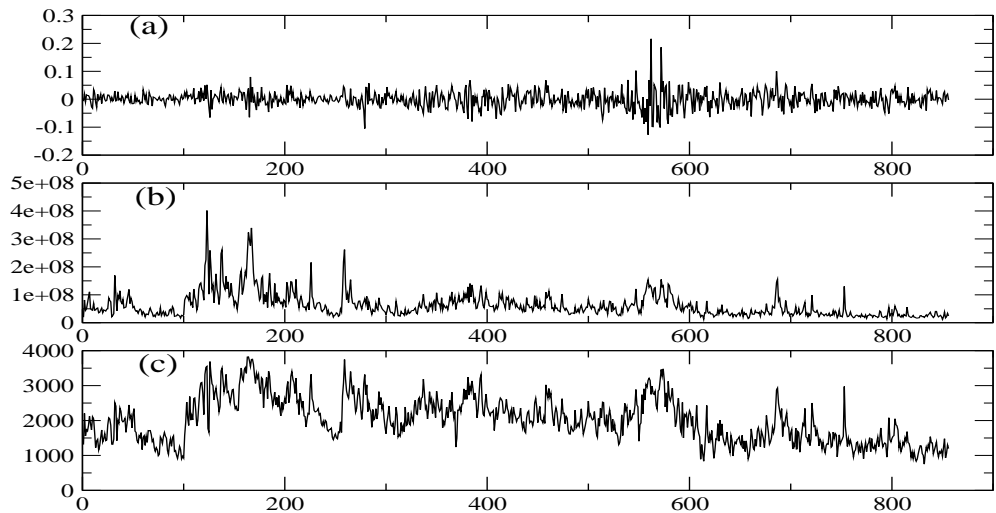}
\caption{
Time series of (a) stock returns, (b) trading volume, and (c) number of transactions for Nippon Steel Co.

}
\vspace{-5mm}
%%\label{fig:corr}
\end{figure}

\section{Empirical Study}

Our analysis is based on four individual stock data, (1) Astellas Pharma Inc., (2) JFE Steel Co., (3) Nippon  Steel Co.,
and (4) Seven \& i Holdings Co., on the Tokyo Stock Exchange.
The sample period of our data  is from  June 3, 2006, to December 30, 2009.
The stock return is defined by the log-price difference: $r_t=100\times (\ln P_t- \ln P_{t-1})$,
where $P_t$ is the closing stock price at day $t$.
Figure~1 shows the time series of (a) returns, (b) trading volume, and (c) number of transactions
for Nippon Steel Co. as a representative case.
We find no strong correlation between the returns and trading volume or number of transactions.
The correlation coefficient $\rho$  between the returns and trading volume (number of transactions) is estimated to be 0.14 (0.02).
On the other hand, we find a strong correlation between trading volume and  number of transactions, $\rho \sim 0.84$.

For parameter estimations, we use the trading volume normalized by its average.
Similarly, we use the number of transactions normalized by its average. 
We perform our GARCH parameter estimation in this study using the MCMC method based on 
the Bayesian inference. The MCMC method we use is the Metropolis--Hastings algorithm
with a multivariate Student's t-proposal density, which has been shown to be particularly efficient
for GARCH parameter estimations \cite{Takaishi1}-\cite{Takaishi6}.
After the first 5000 Monte Carlo samples are discarded as "burn-in" or "thermalization" process, we collect 50000 samples 
for analysis.

Table 1 shows the GARCH parameter results. 
For all stocks, we find that $\alpha+\beta$ is close to 1, implying that
a strong persistence of volatility, in other words, the GARCH effect, exists.

\begin{table}[t]
%\footnotesize
\vspace{-2mm}
\centering
\caption{GARCH parameter results without trading volume and number of transactions.
SD stands for standard deviation.
}
%\label{tb:data1}
%\tabcolsep=35.5pt
%\tabcolsep=19.0pt
\begin{tabular}{ccccc}
\hline
               & $\alpha$ & $\beta$ & $\omega$ &  $\alpha+\beta$  \\
\hline
Astellas Pharma Inc.     & 0.095   & 0.857  & 0.188     & 0.952     \\
 SD                      & 0.018   & 0.026  & 0.066     &      \\
JFE Steel Co.          & 0.096   & 0.895  & 0.126    & 0.991     \\
   SD                  & 0.019   & 0.020  & 0.065    &     \\
Nippon  Steel Co.     & 0.159   & 0.825  & 0.202    & 0.984    \\
     SD               & 0.031   & 0.032  & 0.086    &      \\
Seven \& i Holdings Co.   & 0.117   & 0.873  & 0.0761   &  0.990   \\
       SD                 & 0.027   & 0.028  & 0.0324   &     \\
\hline
\end{tabular}
%\vspace{-2mm}
\end{table}

Table 2(3) shows the GARCH parameter results with trading volume (number of transactions).
We find that even after including the trading volume or number of transactions in the GARCH process,
the value of $\alpha+\beta$ does not change much, except for Astellas Pharma Inc.
Moreover, we find that $\gamma$ is always positive, indicating positive correlations 
between volatility and the trading volume or number of transactions.

\begin{table}
%\footnotesize
%\vspace{-2mm}
\centering
\caption{GARCH parameter results with trading volume. 
}
%\label{tb:data1}
%\tabcolsep=35.5pt
%\tabcolsep=19.0pt
\begin{tabular}{cccccc}
\hline
               & $\alpha$ & $\beta$ & $\omega$ & $\gamma$ & $\alpha+\beta$  \\
\hline
Astellas Pharma Inc.     & 0.232   & 0.251  & 0.043  & 1.97   & 0.483     \\
 SD                      & 0.045   & 0.093  & 0.043  & 0.38   &      \\
JFE Steel Co.            & 0.108   & 0.851  & 0.046  & 0.373  & 0.959     \\
   SD                    & 0.025   & 0.046  & 0.045  & 0.232  &     \\
Nippon  Steel Co.        & 0.177   & 0.790  & 0.102  & 0.241  & 0.967    \\
     SD                  & 0.037   & 0.043  & 0.081  & 0.131  &      \\
Seven \& i Holdings Co.  & 0.166   & 0.786  & 0.0233 & 0.230  &  0.952   \\
       SD                & 0.045   & 0.064  & 0.0257 & 0.136  &     \\
\hline
\end{tabular}
\vspace{-2mm}
\end{table}

\begin{table}[h]
%\footnotesize
\vspace{-4mm}
\centering
\caption{GARCH parameter results with number of transactions.
}
\label{tb:data1}
\begin{tabular}{cccccc}
\hline
               & $\alpha$ & $\beta$ & $\omega$ & $\gamma$ & $\alpha+\beta$  \\
\hline
Astellas Pharma Inc.     & 0.194   & 0.588  & 0.052  & 0.827  & 0.782     \\
 SD                      & 0.049   & 0.127  & 0.054  & 0.370  &      \\
JFE Steel Co.            & 0.100   & 0.872  & 0.050  & 0.233  & 0.972     \\
   SD                    & 0.021   & 0.028  & 0.047  & 0.132  &     \\
Nippon  Steel Co.        & 0.168   & 0.797  & 0.088  & 0.249  & 0.965    \\
     SD                  & 0.034   & 0.041  & 0.079  & 0.150  &      \\
Seven \& i Holdings Co.  & 0.132   & 0.846  & 0.0313 & 0.099  & 0.978   \\
       SD                & 0.033   & 0.037  & 0.0317 & 0.061  &     \\
\hline
\end{tabular}
\vspace{-3mm}
\end{table}

\section{Conclusions}
We examined the relationship between stock volatility and trading volumes or number of transactions
by using four individual stock data of the Tokyo Stock Exchange from  June 3, 2006, to December 30, 2009.
We find that including the trading volume or number of transactions in the GARCH process
does not always reduce the value of $\alpha+\beta$, that is, the magnitude of the GARCH effects.    
Thus, the mixture of the distributions hypothesis using trading volumes or number of transactions as a proxy for the rate of information arrivals is not completely verified.
Since our findings are based on only four individual stock data, it might be interesting to further investigate the robustness of the volatility dynamics using other stock data. 

\section*{Acknowledgment}
Numerical calculations of this work were carried out at the
Yukawa Institute Computer Facility and at the facilities of the Institute of Statistical Mathematics.
%This work was supported by Grant-in-Aid for Scientific Research (C) (No.25330047).
This work was supported by JSPS KAKENHI Grant Number 25330047.

\end{document}